\documentclass[twocolumn]{aastex63}

\usepackage{tabularx}
\usepackage{url}
\usepackage{graphicx}
\usepackage[T1]{fontenc}
\usepackage{ae,aecompl}
\usepackage{booktabs}

\usepackage{blindtext}
\usepackage{longtable}
\usepackage{tabu}

\def\unit #1{\,{\rm #1}}

\newcommand\kms{\rm \,\unit{km\,s^{-1}}}

\newcommand\cmsqi{\rm \,\unit{cm^{-2}}}

\newcommand\kev{\rm \,\unit{keV}}

\newcommand\ergs{\rm \,\unit{erg\,s^{-1}}}

\newcommand\xiunit{\rm \,erg\,cm\,s^{-1}}

\newcommand\lambdaedd{\lambda_{\rm \, Edd}}

\newcommand\msol{M_{\odot}}

\newcommand\nh{\rm N_{H}}

\newcommand\ks{\, \rm ks}

\newcommand\dc{\, \Delta\chi^2}

\newcommand\cd{\,\rm \chi^2/dof}

\newcommand\mpc{\unit{Mpc}}

\newcommand\ev{\unit{\, eV}}

\newcommand\xmm{{\it XMM-Newton}}

\newcommand\suzaku{{\it Suzaku}}

\def\aap{A\&A} \def\apjl{ApJ}  
 \def\apjs{ApJS}

    \submitjournal{ApJ}

\shorttitle{Origin of Reflection features in ESO~511--G030}

\begin{document}

\title{An X-ray spectral study of the origin of reflection features in bare Seyfert 1 galaxy ESO~511--G030}

\author[0000-0003-4790-2653]{Ritesh Ghosh}
\affiliation{Visva-Bharati University, Santiniketan, Bolpur 731235, West Bengal, India.}

\author[0000-0003-2714-0487]{Sibasish Laha} 
\affiliation{Astroparticle physics laboratory, NASA Goddard Space Flight Center,Greenbelt, MD 20771, USA.}
\affiliation{Center for Space Science and Technology, University of Maryland Baltimore County, 1000 Hilltop Circle, Baltimore, MD 21250, USA.}

\correspondingauthor{ritesh ghosh}
\email{ritesh.ghosh1987@gmail.com, riteshghosh.rs@visva-bharati.ac.in}


\begin{abstract}

The reprocessed X-ray emission from Active Galactic Nuclei (AGN) is an important diagnostic tool to study the dynamics and geometry of the matter surrounding the supermassive black holes (SMBHs). We present a broadband (optical-UV to hard X-ray) spectral study of the bare Seyfert 1 galaxy, ESO~511--G030, using multi-epoch \suzaku{} and \xmm{} data from 2012 and 2007 respectively. The broadband spectra of ESO~511--G030 exhibit a UV bump, a prominent soft-excess below $2\kev$, a relatively broad ($\sigma=0.08 - 0.14 \kev$) Fe emission line at $6.4 \kev$ and a weak Compton hump at E$>10\kev$. The soft X-ray excess in ESO~511--G030 can be described either as the thermal Comptonization of disk seed photons by a warm ($0.40^{+0.02}_{-0.02}\kev$), optically thick ($\tau = 12.7^{+0.5}_{-0.4}$) and compact ($< 15 \rm r_{g}$) corona or as the blurred reflection from an untruncated and moderate to highly ionized accretion disk. However, for the blurred reflection, the model requires some extreme configuration of the disk and corona. Both these models prefer a rapidly spinning black hole ($a>0.78$) and a compact corona, indicating a relativistic origin of the broad Fe emission line. We found an inner disk temperature of $\sim 2-3 \ev$ that characterises the UV bump and the SMBH accretes at a sub-Eddington rate ($\lambdaedd = 0.004-0.008$).

 \end{abstract}


\vspace{0.5cm}

\section{INTRODUCTION}

Active Galactic Nuclei (AGNs) are the most powerful emitters in the Universe. The accretion of matter onto supermassive black holes (SMBHs) are the main energy source of this huge emission \citep{1964SPhD....9..195Z, 1969Natur.223..690L}. The accreted matter loses angular momentum and forms a disk-like structure, the simplest theoretical description of which is an optically thick but geometrically thin accretion disk \citep{1973A&A....24..337S}. Thermal emission from this disk mostly emits in the ultraviolet (UV) band, extending to soft X-rays. The seed UV photons are Compton upscattered by an optically thin and hot corona \citep{1973A&A....24..337S, 1993ApJ...413..507H}, around the central SMBH, and produces a power-law spectrum, that dominates the X-ray continuum in AGNs. A part of the coronal emission illuminates the accretion disk and produces reflection features, such as Fe K$_{\alpha}$ emission around $6.4\kev$ and the Compton hump above $10\kev$. Depending on the ionization state of the accretion disk \citep{2005MNRAS.358..211R, 2010ApJ...718..695G} and the proximity of the SMBH, the reflection component can be blurred and distorted by relativistic effects \citep{1991ApJ...376...90L, 2006MNRAS.365.1067C} if originated close to the SMBH; or it can be cold and neutral if reflected farther from the black hole in the outer part of the disk and the torus. The presence of broad and/or narrow Fe emission line in the AGN X-ray spectra generally depicts this idea. In case of Seyfert 1s, apart from the primary powerlaw component and the Fe emission line, a soft-excess component \citep{1985MNRAS.217..105A, 1985ApJ...297..633S} is also observed below $2\kev$. The origin of this excess is still debated. Current studies usually tend to favour either the blurred and ionized disk reflection or the intrinsic disk Comptonization of disk photons. These reflection features originate within a few gravitational radii from the SMBH and hence serve as an important probe of the inner extent of the accretion disk and the black hole spin \citep{2011MNRAS.414.1269W}.

 However, in several cases, these spectral features are modified by the line of sight absorption due to neutral or ionised clouds. Hence, `bare' AGNs, showing no signs of absorption, are crucial to study the emission from the spatially unresolved central region. Notable examples of such sources are Fairall 9 \citep{2001A&A...373..805G, 2009ApJ...703.2171S, 2011MNRAS.415.1895E}, Ark 120 \citep{2004MNRAS.351..193V, 2011MNRAS.410.1251N} and Mrk~509 \citep{2019ApJ...871...88G}. Previous studies of these sources reveal prominent reflection features in the observed X-ray spectra in the form of soft excess and Fe line complex along with the presence of occasional Compton hump. A reflection-based analysis of a sample of 25 bare AGNs by \citep{2013MNRAS.428.2901W} suggests a general preference for rapidly rotating black holes of bare AGNs. In our work, we aim to study the broadband X-ray spectra of a bare AGN to rigorously test the reflection interpretation and subsequently unravel the disk, corona and the central engine properties.

ESO~511--G030 is a bare AGN as have been found by \citet{2010A&A...521A..57T, 2012ApJ...745..107W, 2014MNRAS.441.2613L} and a prime candidate for fulfilling this role due to its prominent reflection features. It is a nearby ($z= 0.0224$) Seyfert 1 galaxy and also one of the X-ray brightest bare Seyferts featured in the Swift 58 month BAT catalogue \citep{2012ApJ...745..107W}. ESO~511--G030 was observed only once by the \xmm{} in 2007 and previous studies revealed the presence of a soft excess and a Fe emission line around $6.4\kev$ \citep{2010A&A...521A..57T, 2012ApJ...745..107W, 2014MNRAS.441.2613L} in the source spectra. \citet{2010A&A...521A..57T, 2014MNRAS.441.2613L} found the presence of a broad Fe K line emission at $6.4\kev$ with an equivalent width of $\sigma=100\pm 15\ev$. Using the same observation, a similar study carried out by \citet{2012ApJ...745..107W}, showed the presence of a narrow Fe emission line ($\sigma = 39^{+8}_{-8}$) in the source spectrum. \citet{2011ApJ...727...19F} studied the source using two \suzaku{} observations and found the presence of narrow Fe K emission lines. They studied the Fe line flux variability of a sample of sources including ESO~511--G030 and suggested that X-ray reflection by a torus is the main origin of this narrow Fe K emission. Previous studies of ESO~511-G030 mentioned here have primarily focused on understanding the details of the warm absorber and high-velocity outflows. In this work we address the following science questions: 1. The nature and origin of the Fe line emission in the source spectra 2. Detection or nondetection of the soft excess and the Compton hump and finally 3. If we do detect a broad Fe line, can we constrain the black hole spin? We used the multi-epoch broadband ($0.001- 50\kev$) spectra of ESO~511-G030 using the \xmm{} and \suzaku{} observations. We have also used OM data from \xmm{} to constrain the UV bump, necessary to investigate the accretion disk and black hole properties with physically motivated models. Throughout this paper, we assumed a cosmology with $H_{0} = 71\kms \mpc^{-1}, \Omega_{\Lambda} = 0.73$ and $\Omega_{M} = 0.27$.
The paper is organised as follows: Section \ref{sec:obs} describes the observation and data reduction techniques. The steps taken in the spectral analysis are discussed in Section \ref{sec:analysis}. Section \ref{sec:results} discusses the results followed by conclusions in Section \ref{sec:conclusions}.    \\

\section{Observation and data reduction}\label{sec:obs}

ESO~511--G030 was observed only once by \xmm{} in 2007 August 5 and twice by \suzaku{} in 2012 July 22 and in 2012 August 17. The details of the observations and the short notation of the observation ids are mentioned in Table \ref{Table:obs}. We processed the EPIC-pn and OM data using V18.0.0 of the Science Analysis Software (SAS)~\citep{2004ASPC..314..759G} with the task {\it epchain} and {\it omichain} respectively and filtered them using the standard filtering criterion. We used the latest calibration database available at that time. The EPIC-pn data was preferred due to its higher signal to noise ratio compared to MOS. We checked the background rate above $10\kev$ for any flaring particle background and used a rate cutoff of $< 1 \, \rm ct s^{-1}$ to create the good time intervals. To extract the source spectrum and light curve, we selected a circular region of 40 arcsec, centred on the centroid of the source. For the background spectrum and light curve, we chose nearby circular regions, located on the same CCD, that are free of any sources. The EVSELECT task was used to select single and double events for EPIC-pn ($PATTERN<=4$, $FLAG==0$) source event lists. We created the time-averaged source + background, the background spectra and the corresponding response matrix function (RMF) and auxiliary response function (ARF) for each observation using the {\it xmmselect} command in SAS. We checked for pile up in the \xmm{} source spectra using the command {\it epatplot} and found that observation (obs1 from now on) is not piled up. The \xmm{} spectra were grouped by a minimum of 50 counts per channel and a maximum of three resolution elements using the command {\it specgroup}. For the OM data, we obtained the count rates in four active filters (B, UVW1, UVM2 and UVW2) by specifying the RA and Dec of the source in the source list file obtained by the {\it omichain} task. The {\it om2pha} task was used to produce the necessary files for OM photometric data to be analysed simultaneously with EPIC-pn in {\sc XSPEC}.

 ESO~511--G030 was observed twice by \suzaku{} on 22nd July and 17th August 2012 (See Table \ref{Table:obs}). The three X-ray Imaging Spectrometer (XIS) \citep{2007PASJ...59S..23K} along with the Hard X-ray Detector (HXD) \citep{2007PASJ...59S..35T} cover the broad energy band of $0.2-50 \kev$. In both observations (from now on obs2 and obs3 respectively), all the data were obtained in standard XIS ($3 \times 3$ and $5 \times 5$) and HXD data modes. The data reduction technique of XIS and HXD-PIN is the same as \citet{2018MNRAS.479.2464G}, following the standard \suzaku{} data reduction. We used {\sc HEASOFT(v6.27.2; HEASARC 2014)}, software and the recent calibration files to reprocess the \suzaku{} data. For the non-imaging HXD/PIN data, we used the appropriate tuned background files provided by the \suzaku{} team and available at the {\sc HEASARC} website. The XIS front-illuminated  spectra were coadded to enhance the signal to noise ratio. We grouped the XIS spectral data off to a minimum of 200 and 100 counts in each energy bin for obs1 and obs2 respectively. We also grouped the PIN data using the command {\it grppha} in the {\sc HEASOFT} software to produce $\sim 30$ energy bins with more than 20 counts per bin in the source spectra.

\section{Data Analysis and spectral fitting}\label{sec:analysis}

We used {\sc XSPEC} \citep{1996aspc..101...17a} version 12.11.0m to analyse all the data sets and all errors quoted on the fitted parameters reflect the 90 per cent confidence interval corresponding to $\dc =2.7$ \citep{1976ApJ...208..177L}. We excluded the XIS data in the $1.7 - 2.3\kev$ from our spectral analysis due to calibration uncertainties. We used {\it tbabs} model to estimate the effect of Galactic absorption and set the scattering cross-section to Verns and abundances to Wilms values. In all our spectral fitting, we adopted the Galactic column density value of $\rm N_{H}= 4.34\times 10^{20} \rm cm^{-2}$ \citep{1990ARA&A..28..215D}.

\begin{figure*}
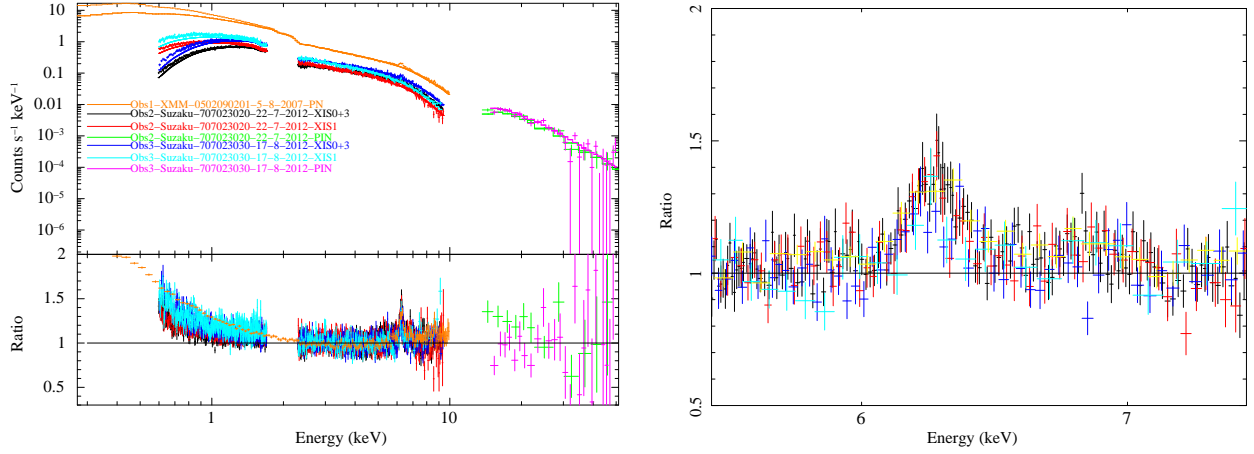

  \centering 

\hbox{
\includegraphics[width=6.0cm,angle=-90]{Best_fit_excess_all_data_ratio.ps}
\includegraphics[width=6.0cm,angle=-90]{excess.abs.po.all.spectra.RG.only.ratio.zoomed.eso511mg030.ps} 
}\caption{ {\it Left:} The $4.0-5.0\kev$ \suzaku{} and \xmm{} spectra of ESO~511--G030 fitted with an absorbed powerlaw and the rest of the $0.3-50.0\kev$ dataset extrapolated. Bottom panel: The broadband residuals from the fit above, showing the presence of soft X-ray excess, an Fe line complex and a hard X-ray excess (at $E>10\kev$). The X-axis represents observed frame energy. {\it Right} A zoomed in version of the residual at the $6\kev$ region. } \label{fig:check_excess}

\end{figure*}

We started our analysis with a preliminary look at the spectra of the source. We used an absorbed power-law model in the $4-5\kev$ energy band and when  extrapolated to the rest of the X-ray band revealed a prominent soft excess (below $2\kev$), a Fe line complex at $6-7\kev$ for all three observations. We did not find any significant excess above $10\kev$ in any of the two \suzaku{} HXD-PIN data sets (See Fig.~\ref{fig:check_excess}). To address our science goals, we organize the spectral analysis of multi-epoch broadband data, into three parts: above $3\kev$, the full X-ray band and the joint X-ray and UV. Here, the detailed analysis of the above $3\kev$ energy band will help us reveal the true nature of the Fe emission line. On the other hand, the study of full X-ray and joint X-ray plus UV data of ESO~511--G030, with physically motivated models, will help us investigate the nature of accretion disk and black hole properties and the role they play in the origin of the soft excess. The latest XSPEC version used here provides us with the best-fit statistics of individual data sets during simultaneous model fitting. Hence, a simultaneous multi-epoch spectral fit of \xmm{} and \suzaku{} spectra was carried out in all three cases.

\subsection{Investigating the Fe emission line}\label{subsec:3kev}

To investigate the excess emission in the $6-7\kev$, we first introduced a narrow Gaussian to the absorbed powerlaw model. An energy-independent multiplicative factor was used to account for the relative normalization between different instruments of \suzaku{}. Assuming a bare nature of the source we did not include any neutral or partially ionised absorber model to the set of models used here. In {\it XSPEC} the model reads as {\tt constant$\times$ tbabs $\times$ (po$+$ zgauss)}. The best-fit statistics for this set of models is $\cd=1587/1414$. The line energy is consistent with $6.4\kev$ for all three observations and the equivalent width ranges between $62\ev$ to $94\ev$. We found the source frame line width $\sigma = 0.08-0.14\kev$ between observations, which indicates the presence of a broad Fe emission line in the source spectra. The broad Gaussian if replaced with a narrow Gaussian line ($\sigma = 0.01\kev$) provides a relatively poor fit statistics to the data sets ($\cd = 1603/1417$). Next, we replaced the broad Gaussian line with a more realistic broad Fe line profile {\it diskline}, \citep{1989MNRAS.238..729F}, which resulted in a similar fit statistics ($\cd = 1587/1414$). In the {\it diskline} model, the inner and outer radius of the disk was fixed at $r_{\rm in}= 6r_{\rm g}$ and $r_{\rm out}= 400r_{\rm g}$ respectively for all three observations. The power law dependence of emissivity ($\beta$), the line energy and the model normalizations are made free for obs 1 and obs2. The inclination angle is tied between observaions. The best-fit line energy is consistent between observations with a value of 6.42 keV. The high value of emissivity profile ($\rm q=0.9-9.1$) indicates a large amount of coronal radiation intercepts the disk. 

Alternatively, a partial covering absorber model (neutral or ionized) may mimic the red wing of the broad Fe emission line in the broad-band X-ray spectrum. The partial covering model with neutral or ionized absorption ({\it zpcfabs} or {\it zxipcf}) when multiplied with the absorbed (Galactic) power-law model resulted in a poor statistics for all the three data sets ($\cd =1868/1421$ and $\cd = 1873/1422$ for {\it zpcfabs} or {\it zxipcf} respectively). See Table \ref{Table:above3kev} for the comparison of a different set of models used to fit the Fe line emission. Our analysis indicates the presence of a broad Fe K$\alpha$ emission line in the X-ray spectra of all three observations. Next, we study the full X-ray band to further characterize the reflection features. 

\subsection{The full X-ray band}\label{subsec:X-ray}

We initially analysed the broadband X-ray data with a phenomenological set of models to test the presence and strength of reflection features in the source spectra. The baseline phenomenological model used here includes a neutral Galactic absorption (tbabs), a multiple blackbody component to model the soft excess \citep[diskbb,][]{1984PASJ...36..741M} and the coronal emission described by a power-law. In addition, we used a {\it diskline} model to describe excess emission at around $6.4\kev$. An energy-independent multiplicative factor was introduced to account for the relative normalization of different XIS and PIN instruments. In {\it XSPEC} notation, the phenomenological model reads as {\tt constant$\times$ tbabs$\times$ (powerlaw$+$diskbb$+$diskline)}. Addition of a {\it ztbabs} model did not improve the fit statistics and we conclude that all three observations are free from any intrinsic neutral absorption of the host galaxy. We included the {\it pexrav} model to check for the reflection component from the neutral medium which contributes to the Compton hump. The \xmm{} EPIC-pn data ($< 10 \kev$) is insufficient to constrain the {\it pexrav} reflection parameters and hence tied with the \suzaku{} observation (obs2) with longer exposure. We found a significant improvement in the fit statistics upon the addition of {\it pexrav} ($\dc \sim 72$ for 4 dof) indicating the presence of neutral reflection component in the hard X-ray band above $10\kev$. The phenomenological model provides a satisfactory description to the two \suzaku{} as well as \xmm{} spectra. The best-fit parameter values obtained using the phenomenological models are marginally consistent between observations and quoted in Table \ref{Table:pheno}. We also quoted the improvement in statistics ($\rm \dc/dof$) for each spectrum upon the addition of different model components to determine the statistical significance of the model (See Table~\ref{Table:pheno}). The line energy of the Fe emission is consistent with $6.4\kev$ for all three observations. Our results indicate that a prominent soft excess and a broad Fe K$_{\alpha}$ emission line is present in all the observations that are usually considered as signs of relativistic reflection from the disk.

However, the origin of soft excess is still debated and apart from the relativistic reflection from disk, intrinsic disk Comptonization has also been found to be a viable physical model in other Seyfert 1 galaxies. In our work, we used two sets of physical models to describe both the soft excess and the Fe emission line in the \suzaku{} and \xmm{} X-ray spectra of ESO~511--G030. In {\it XSPEC}, our first set of physical models reads as {\tt constant $\times$ tbabs $\times$ (relxill$+$MyTorus)}. Here, the {\it relxill} model, version 1.3.7, \citep{2014ApJ...782...76G} describes the soft X-ray excess, the power-law continuum, the broad Fe K$\alpha$ emission line and the reflection of primary hard X-ray photons off an ionized accretion disk, on the other hand, the Compton reflection of hard X-ray photons off cold, neutral material is modelled with {\it MYTorus}. The {\it MYTorus} model comprised of three components: first, the torus-absorbed primary power-law, second, the scattered emission (MYTorusS) due to the reflection of primary hard X-ray photons from the torus and third, the iron FeK$_{\alpha}$ and K${\beta}$ lines (MYTorusL), which are assumed to arise due to the reflection by the torus. ESO~511--G030 is a type-1 AGN hence we do not expect any obscuration due to torus along the line of sight. We have used only MYTorusS and the MYTorusL component in our spectral fitting and tied their column densities together. The column density of {\it MYTorus} gives a value in an equatorial direction of a dusty torus, so a Compton-thick value is expected even in type 1 sources \citep{2016MNRAS.462.4038Y, 2019MNRAS.486.3124L, 2020MNRAS.tmp.2348G}. When made free, the {\it MYTorus} column density pegged at $10^{25}\rm cm^{-2}$ and we got a lower limit of $2.9\times 10^{24}\rm cm^{-2}$. Hence, we froze the {\it MYTorus} column density to a value of $10^{25}\rm cm^{-2}$, assuming that a Compton thick torus is responsible for this emission. We also fixed the inclination angle between the torus polar axis and the observer's line of sight to 45 degrees for both sets of physical models. The photon index of the {\it MYTorus} model was tied with that of the photon index of the primary continuum of {\it relxill} model. In the {\it relxill} model, the primary hard X-ray emission from corona illuminates the accretion disk and produces fluorescence emission lines. These fluorescence lines get blurred and distorted due to extreme gravity near the central supermassive black hole and produce the soft excess. The transition between Newtonian and relativistic geometry is marked by a breaking radius $\rm r_{br}$. In our work, the emissivity index of reflection from the disk outside $\rm r_{br}$ (q2) is fixed at 3, as in Newtonian geometry, for a point source, the emissivity at a large radius out from the source has a form $\rm r^{-3}$. The emissivity index inside $\rm r_{br}$ falls under the relativistic high-gravity regime and calculations \citep{2001MNRAS.321..605D, 2003MNRAS.344L..22M, 2011MNRAS.414.1269W} suggest a very steeply falling profile in the inner regions of the disk. Hence, we made the emissivity index of the inner part (q1) of the accretion disk inside $\rm r_{br}$, free to vary in between $3 < \rm q1 < 10$.

 During our simultaneous analysis of joint \xmm{} and \suzaku{} data sets, the parameters that are unlikely to change within human timescale, e.g., the black hole spin and the inclination angle, were tied between all observations. Apart from the power-law photon index, normalization and the reflection fraction (R), all other parameters were tied between the two \suzaku{} observations. The {\it MYTorus} model requires a hard X-ray spectrum to constrain its parameters. Due to the absence of $> 10\kev$ spectrum in \xmm{} observation, the {\it MYTorus} parameters were tied with best-fit parameter values of obs2. This set of physical models produced a satisfactory fit with fit statistics $\cd= 3431/2981$ however we observed some excess in the residual at $\sim 0.5$ and $0.9\kev$. Addition of two narrow Gaussian line profile to the model yielded an improved fit ($\cd = 3345/2976$). We estimated the high energy cutoff of the primary power-law component using the \suzaku{} observation with the longest exposure (obs1). We found a lower limit to the high energy cutoff with $\rm E_{c} >297\kev$. We tied this value for all other observations. A rapidly spinning black hole ($>0.98$) is required by the model to fit the X-ray spectra. We carried out a test to find out if the black hole is indeed maximally spinning. First, we froze the spin parameter to zero and kept the inner radius ($\rm r_{in}$) of the {\it relxill} model fixed to the inner circular stable orbit for a non-rotating black hole ($\rm r_{in} = 6\rm r_{g}$) and fitted the data sets. Next, we froze spin to 0.998 and fixed the $\rm r_{in}$ value to that of a maximally spinning black hole ($\rm r_{in} = 1.24\rm r_{g}$). We found that the maximally spinning scenario provides a significantly better fit statistics ($\dc = 501$). Following this result, we fixed the inner radius we fixed the inner radius of the disk at $\rm r_{in} = 1.24\rm r_{g}$ throughout the rest of the fit and allowed the spin parameter to vary freely. We found the iron abundance of the material in the accretion disk to be significantly lower than the solar value ($\rm A_{Fe} <0.7$) for the two \suzaku{} observations. The best-fit reflection fraction (R) ranges between $\rm R= 0.5-5.3$ for the three observations. The best-fitting model parameters along with the best-fitting statistics for each observation are quoted in Table \ref{Table:X-ray}. 

We have also tested other flavours of {\it relxill}, such as {\it relxillD} and {\it relxilllP}. The model {\it relxillD} allows a higher density for the accretion disk (between $\log N/\rm cm^{-3} =15.3$ to $\log N/\rm cm^{-3} =18 $) and the model {\it relxilllP} assumes a lamp post geometry and determines the variation in the position of the hard X-ray emitter. Both these models produced a relatively poor statistics ($\cd = 3434/2973$ for {\it relxillD} and $\cd = 3469/2976$ for {\it relxilllP} respectively) compared to {\it relxill}, for all the observations and no significant variation either in the disk density or in the position of the hard X-ray emitter was found. 

For the second set of physical models, we use intrinsic thermal Comptonization from a warm corona ({\it optxagnf}). In the {\it optxagnf} model, thermal comptonization of disk photons by a warm ($\rm T\sim 0.5-1\kev$) and optically thick ($\tau\sim 10-20$) corona \citep{1998MNRAS.301..179M, 2012mnras.420.1848d}, that lies in the inner part ($10-20\rm r_{g}$) of the accretion disk, produces the soft excess. The gravitational energy released in the accretion process fuels the disk emission in UV, the soft X-ray excess and the power-law emission. The model normalization flux is determined by the source Eddington rate, the black hole mass, the black hole spin and the luminosity distance. Hence in our analysis, we froze this value at unity. The hard excess emission, in this set of models, arises exclusively due to neutral Compton reflection from the torus. In XSPEC notation, the model reads as {\it(constant $\times$ tbabs$\times$(optxagnf + MYTorus))}. We used a black hole mass of $\sim 4.57\times 10^{8}\msol$, adopted from \citet{2012A&A...542A..83P}, who derived it using X-ray variability of the source. In our simultaneous fit of \suzaku{} and \xmm{}, apart from the black hole spin, all other {\it optxagnf} model parameters were made free to vary. On the other hand, for the two \suzaku{} observations, except for the Eddington rate, photon index and the parameter $\rm f_{pl}$, that determines the fraction of powerlaw emitted as soft excess, all other parameters were tied between observations. This set of physical models produced a poor fit statistics ($\cd = 3411/2992$) compared to our first set of models, however, we observed similar excess emission in the residual at $\sim 0.5\kev$ and $\sim 0.9\kev$. Addition of two Gaussian line profile to the set of models provided similar fit statistics ($\cd = 3351/2987$) compared to the reflection model {\it relxill}. All three observation show sub-Eddington accretion rate ($\sim 0.004-0.008$) and infer a maximally spinning black hole ($> 0.98$). Table \ref{Table:sed} includes the best-fitting parameters obtained using this model. Next, we included the OM data from the \xmm{} and tried to get a better constrain on the {\it optxagnf} model parameters.

\subsection{The joint UV and X-ray band}\label{subsec:broadband}

The thermal Comptonization model uses disk photons to produce the powerlaw spectra and the soft excess. Therefore, the UV bump is required to constrain the {\it optxagnf} parameters. In this section, we re-fit the X-ray data along with the simultaneously obtained UV spectra to constrain the model parameters.

We used the optical/UV fluxes in four bands (B, W1, W2 and M2) measured with the \xmm{} to constrain the thermal emission from the disk. The Galactic extinction correction was done following \citep{1999PASP..111...63F} reddening law with $\rm R_{v} = 3.1$ and is taken into account by the {\it REDDEN} model. The parameter value used here is $\rm E_{(B-V)}= 0.056$ \citep{2011apj...737..103s}. We have added 5 per cent systematic error to the OM fluxes to account for the host galaxy contamination, the nuclear emission lines and the intrinsic reddening, which produce a considerable amount of systematic uncertainty in the measured optical-UV continuum flux. The two sets of physical models, {\it relxill} and {\it optxagnf}, were used to jointly fit the OM and EPIC-pn data of \xmm{} along with the two \suzaku{} observations. The best-fitting parameter values are quoted in Table \ref{Table:sed}. The model parameters of {\it relxill} and {\it MyTorus} were frozen in the joint fit as they were determined by the X-ray energy band only and we added a {\it diskbb} model to fit the OM data. The model resulted in a fit-statistic of $\cd = 3349/3001$. In the case of {\it optxagnf} model, we froze the {\it MyTorus} model parameters and allowed {\it optxagnf} model parameters to vary freely. We find that this second set of physical models yield a marginally poor fit statistics ($\cd = 3369/2989$) (See Table~\ref{Table:sed}) compared to the ionized reflection model nevertheless both provide a satisfactory description to the broadband optical-UV to hard X-ray data. 

\begin{figure*}
  \centering 

\hbox{
\includegraphics[width=8.0cm,height=6.5cm,angle=0]{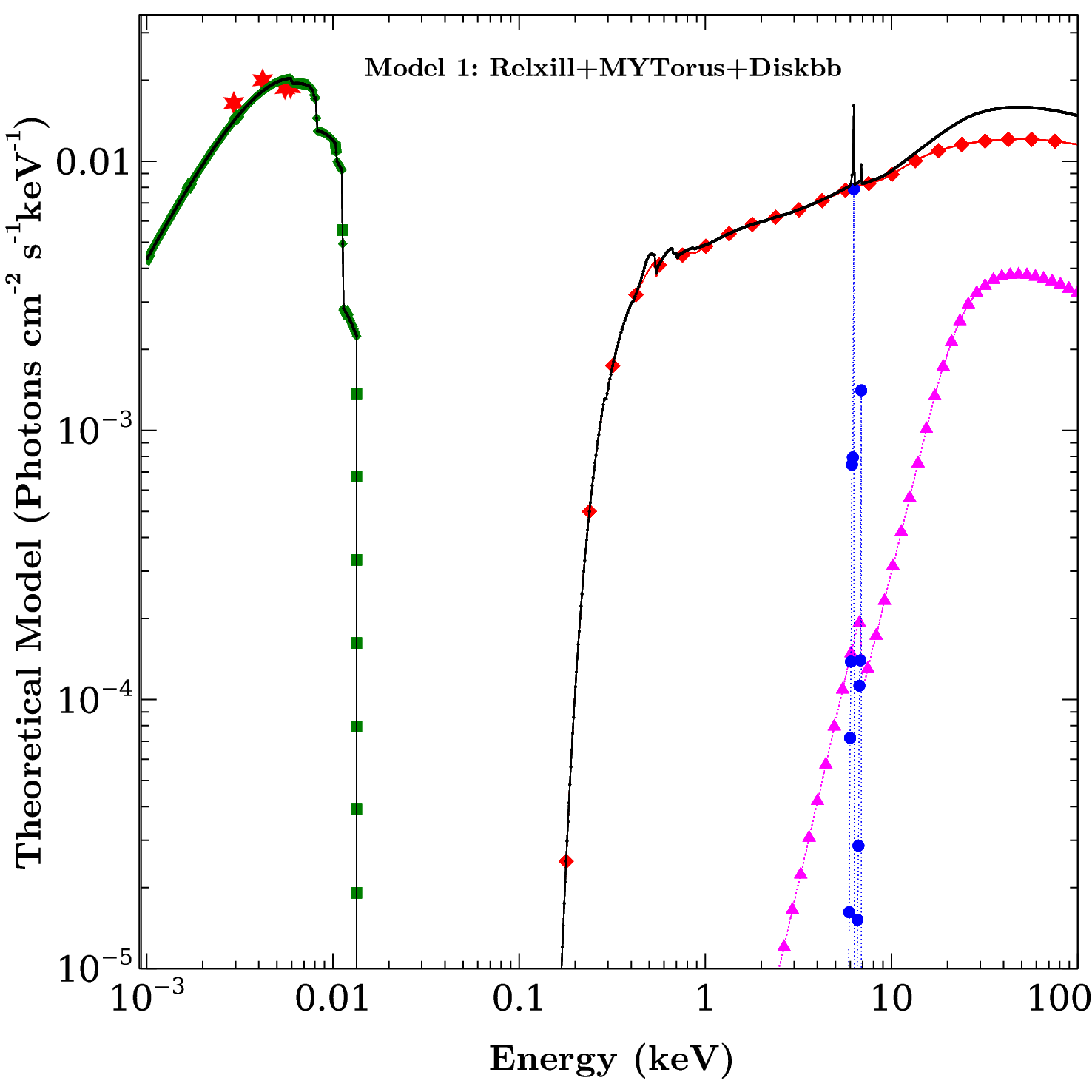}
\includegraphics[width=8.0cm,height=6.5cm,angle=0]{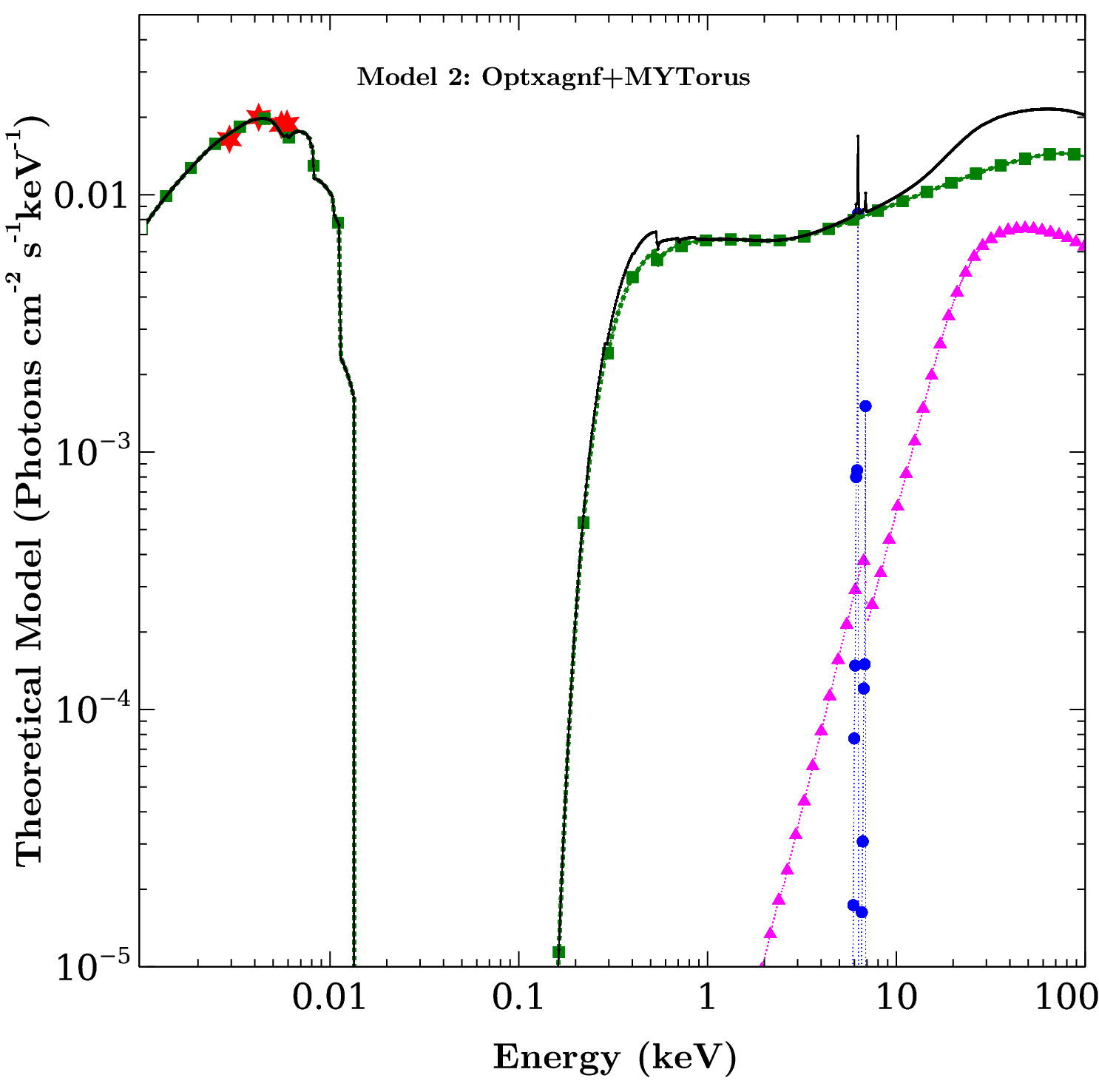} 
}\caption{ {\it Left:} Left. The theoretical models are plotted after the joint fitting of broadband optical-UV to hard X-ray spectra of \suzaku{} and \xmm{} of the source ESO~511--G030 with the {\it relxill} (red-diamond) and {\it MYTorus} (line component in blue-circle and scattered component in magenta-triangle) plus the {\it diskbb} (in green-square) model. {\it Right:} Same for the second set of physical models: thermal Comptonization model {\it optxagnf} (in green-square) plus {\it MYTorus} (line component in blue-circle and scattered component in magenta-triangle). The OM data points are plotted as red-star. The X-axis represents the observed frame energy. } \label{fig:sed}

\end{figure*}

Our simultaneous analysis of UV and X-ray data enabled us to get a better constrain on the warm corona properties, e.g., the electron temperature, the radius and the opacity(See Table~\ref{Table:sed}). Our results suggest the presence of a warm ($\rm kT_{e} = 0.1-0.5\kev$), optically thick ($\tau = 8-13$) and compact ($\rm r_{corona} = 5.9-11.8 \rm r_{g}$) corona and a maximally rotating ($a>0.98$) black hole. The best-fitting models obtained from the joint fitting of \xmm{} and \suzaku{} are shown in Fig.~\ref{fig:sed}. A detailed discussion of these results is followed in the next section.

\section{Results and Discussion}\label{sec:results}

We have extensively studied the broadband optical-UV to hard X-ray ($0.001-50\kev$) spectra of Seyfert 1 galaxy ESO~511--G030 using all the archival data from \suzaku{} and \xmm{}. In our work, for the first time, we have used physically motivated models to describe the observed reflection features in the source spectra. Our study enabled us to unambiguously constrain the accretion disk properties and subsequently the black hole spin. We find that both reflections from an ionized disk and intrinsic thermal Comptonization describe the soft excess well. 

However, the thermal Comptonization model yielded a much better fit statistics to the optical-UV to hard X-ray source spectra. We measured the $2-10\kev$ unabsorbed luminosity ($\log \rm L_{2-10\kev}$) of the source and found it to be consistent between 2007 and 2012. The observed values are $43.52\ergs$, $43.50\ergs$ and $43.66\ergs$ for obs1, obs2 and obs3 respectively. We estimated of the bolometric luminosity of ESO~511--G030 ($4.76\times 10^{43}\ergs$), using the following equation: $\log \kappa_{\rm Lbol}=1.561-1.853\times \alpha_{\rm OX} + 1.226 \times \alpha_{\rm OX}^2$ \citep{2010A&A...512A..34L, 2018MNRAS.480.1522L}. We used the OM flux value of \xmm{} observation, the value of $\rm L_{2-10\kev}$ and the $\alpha_{\rm OX}$, where $\alpha_{\rm OX}$ is the power-law slope joining the $2 \kev$ and the $2500 \rm \AA$ flux. ESO~511--G030 has a black hole mass of $\sim 4.57\times 10^{8}\msol$ \citep{2012A&A...542A..83P} which implies an Eddington ratio of $0.002$. This result is in contrast with \citep{2018MNRAS.480.3898N}, who studied the changing look AGN MRK 1018 with optxagnf and fixed the black hole spin value to zero. But in our case, we found a model-dependent high black hole spin and a strong soft X-ray excess that is present even for a small fraction (0.4-0.8\%) of $\lambdaedd$. Our measured Eddington rate is consistent with the value previously found by \cite{2014MNRAS.441.2613L}, indicating a steady accretion state in the source during  2007 and 2012. Using the reflection model we got a lower limit to the high energy cut off value ($\rm E_{c} >297\kev$) of the primary power-law component. Below we discuss the main results investigating the spectral features observed in this source.

\subsection{The Fe K$_{\alpha}$ line and the Compton hump}\label{subsec:Fe.line}

\begin{figure*}
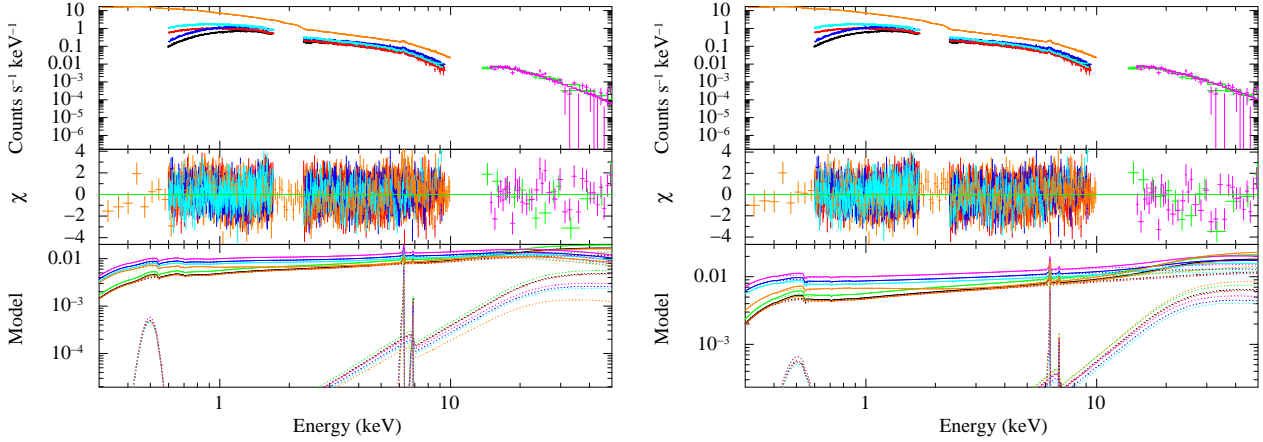

  \centering 

\hbox{
\includegraphics[width=6.0cm,angle=-90]{Best_fit_absorbed_relxill_mytorus_withspin.all.spectra.eso511mg030_zgauss.ps}
\includegraphics[width=6.0cm,angle=-90]{Best_fit_absorbed_optxagnf_mytorus_withspin_zgauss.all.spectra.eso511mg030.ps} 
}\caption{ {\it Left:} Left. The $0.3-50.0\kev$ \suzaku{} and \xmm{} spectra of the source ESO~511--G030 with the best-fitting reflection model and residuals. The relxill model describing simultaneously the soft X-ray excess, the broad Fe $\rm K_{\alpha}$ emission line, and the relativistic reflection hump in the hard X-rays, the MYTorus model describing the narrow Fe $\rm K_{\alpha}$ and Ni emission lines along with the Compton hump due to distant neutral reflection, are plotted in the lower panel. {\it Right:} Same for the Comptonization model optxagnf. The X-axis represents the observed frame energy. } \label{fig:phys}

\end{figure*}

Previous studies of the X-ray spectra of ESO~511--G030 have mostly revealed the presence of a relativistically broad Fe emission line. \citet{2010A&A...521A..57T,2014MNRAS.441.2613L} studied the 2007 \xmm{} data and found a broad Fe emission line at $6.4\kev$ with an equivalent width of $\sigma=100\pm 15\ev$. On the other hand, using the same observation, \citet{2012ApJ...745..107W} studied the X-ray broadband properties and detected the presence of a narrow Fe emission line ($\sigma = 39^{+8}_{-8}$) in the source spectrum. \citet{2011ApJ...727...19F} studied the source using two \suzaku{} observations and found the presence of narrow Fe K lines. In our work, we have investigated the above $3\kev$ energy band with a set of models to unambiguously determine the nature of the Fe emission line. We found the presence of a broad Fe emission line at $\sim 6.4\kev$ in all the spectra of ESO~511--G030. The best-fit Fe emission line $\sigma$ ($0.08-0.14\kev$) and equivalent width ($62-94\ev$) are significantly broad than the equivalent width of typical narrow Fe emission line found in nearby Seyfert 1 galaxies. The broad Gaussian line profile and the {\it diskline} model provides a better description of the Fe line emission than other sets of models and indicates the presence of a relativistically-broadened Fe emission line from an accretion disk around a rotating black hole. Our broadband spectral modelling of ESO~511-G030 with both set of physical models favours a rapidly spinning black hole and supports this idea. In all the observations, the Fe line centroid energy was consistent with $6.4\kev$, which indicates that the iron is neutral or in low ionization state. The best-fit ionization parameter value of the \xmm{} observation ($\log\xi =2.3^{+0.1}_{-0.1}$) supports this idea. However, for the two \suzaku{} observations, we got a highly ionized accretion disk ($\log\xi =3.2^{+0.1}_{-0.1}$) compared to obs1. 
From Fig.~\ref{fig:check_excess} we note that the Compton hump above $10\kev$ is relatively weak but the addition of a {\it pexrav} model to the phenomenological set of models did improve the fit-statistics ($\dc \sim 72$ for 4 dof) significantly. From Fig.~\ref{fig:phys} we find that both neutral and ionised reflection component provides a good description of the Compton hump. With the current data quality in \suzaku{}, we are unable to comprehensively detect or separate out the contributions of the ionized and neutral reflection components and further deep observations of Seyfert 1s are required.

\subsection{The soft excess}\label{subsec:soft.excess}

Our analysis of the full X-ray band of ESO~511--G030 revealed the presence of a prominent soft excess below $2\kev$ for all three observations. We investigated the broadband multi-epoch data in detail to identify the disk and black hole properties responsible for this excess emission. Recent studies suggest that both relativistic reflection from an ionized accretion disk and the intrinsic thermal Comptonization of disk photons can successfully describe the soft excess in Seyfert 1s \citep{2020MNRAS.tmp.2348G, 2019ApJ...871...88G, 2019MNRAS.489.5398W, 2018MNRAS.478.4214E}. These two models assume two very different geometries and physical properties of the accretion disk and the corona (e.g., the position, temperature and the opacity). However, it is not easy to distinguish them on statistical grounds alone and often extreme values of certain model parameters are used to favour one model over the other. In some nearby Seyfert 1s, e.g., HE~1143-1810 \citep{2020A&A...634A..92U} and Zw~229.015 \citep{2019MNRAS.488.4831T}, the ionized disk reflection model was ruled out due to the relatively extreme values of the iron abundance, the inclination angle and reflection fraction, compared to other Seyfert 1s. On the other hand, in Mrk~478 \citep{2019MNRAS.489.5398W}, the flux variability between data sets was better described by ionised disk reflection, compared to the thermal Comptonization model.  

 In our work, we have studied the reflection features observed in the ESO~511--G030 spectra and got comparable fit statistics for both of these physical models. The soft excess is equally well described by both these models. The best-fit parameter values obtained for the {\it optxagnf} model were in a range detected in typical Seyfert 1 galaxies (See Table \ref{Table:X-ray}) although, we were unable to constrain the optical depth in obs2 (See Table \ref{Table:sed}). Similarly, for the reflection model, some model parameters require extreme values to model the soft excess and the X-ray energy band. The best-fit photon index $\Gamma$ of the model {\it relxill} ranges between $1.7 - 2.0$ for all three observations. The best-fit ionization parameter value of the model ($\log \xi \sim 2-3 \xiunit$) suggests a transition between moderate to highly ionized disk between obs1 and obs2 respectively. In the reflection model {\it relxill}, the black hole spin and the inner radius are degenerate and hence fixing the inner radius to the ISCO provides a better constrain on the spin. Hence we fixed the inner radius to $1.24 \rm r_{g}$ and obtained a maximally spinning black hole ($\rm a>0.98$). We were able to constrain the inclination angle parameter and found a best-fit value of $27^{+3}_{-1}$. The iron abundance of the reflecting medium ($\rm A_{Fe}$) or the disk is marginally consistent with the solar abundance value for the \xmm{} observation however when made free we got an upper limit of $<0.6$ for the two \suzaku{} observations. We also found a large variation in the best-fit value of reflection fraction (R) between the three observations, ranging between 0.5 to 5.3, although consistent with other Seyfert 1s. This value R, determines the ratio of photons entering the disk to the ones escaping to infinity and indicate that most of the hard X-ray photons are entering the disk and results in a highly ionized disk. These results along with the high emissivity index ranging between $4.6-8.2$ imply that major part of the soft excess has originated from a region very close to the central supermassive black hole due to reflection of hard X-ray photons from a highly ionized untruncated accretion disk. Although very high values of black hole spin and the reflection fraction or lower iron abundances in the disk are occasionally reported for AGNs, a corona placed so close to the black hole implies a very extreme configuration of the accretion disk and the corona. 

On the other hand, the intrinsic thermal Comptonization model {\it optxagnf} combined with the neutral reflection from the torus, modelled by {\it MyTorus}, provides a similar fit statistics compared to the ionized reflection. Most of the best-fit parameter values obtained from the fitting of the full X-ray band are well constrained and are typical of other Seyfert 1 galaxies \citep{2020MNRAS.tmp.2348G, 2018A&A...609A..42P, 2020A&A...634A..92U}. In this model, the soft excess originates due to Comptonization of thermal disk photons by a warm and optically thick corona that covers roughly $5-12 \rm r_{g}$ of the inner accretion disk \citep{2013A&A...549A..73P}. However, we note that the accretion rate in {\it optxagnf} is also determined by the outer accretion flow beyond $\rm r_{corona}$ that contributes primarily in the optical-UV band. This explains our inability to constrain some of the model parameters e.g., the $\rm r_{corona}$ and the black hole spin. Hence we discuss the best fit parameter values obtained from the simultaneous analysis of the optical-UV to hard X-ray broadband data with {\it optxagnf} as they provide us with a more clear view of the accretion disk and the black hole properties.


\subsection{The optical-UV to hard X-ray continuum}\label{subsec:sed}

In the {\it optxagnf} model, the gravitational energy released in accretion, powers three distinct emission components - the UV bump, the soft excess, and the hard X-ray power law. As expected, {\it optxagnf} combined with neutral reflection from torus yielded better fit statistics (See Table~\ref{Table:sed}) compared to the ionized reflection model. From the joint fits of all three observations, using the {\it relxill$+$diskbb} model, we find that the disk blackbody has a much colder temperature of $\rm kT = 3.8^{+0.7}_{-1.2} \ev$. This value is comparable with the inner radius temperature, $\sim 2 \ev$, of the accretion disk that is accreting at a rate of $\lambdaedd = 0.004$, obtained for obs1, using {\it optxagnf} model. This best-fitting accretion rate is close to our estimated value ($\lambdaedd = 0.002$) obtained using the bolometric luminosity of ESO~511--G030. The soft excess in \xmm{} observation is well described as a warm ($\rm kT = 0.4^{+0.02}_{-0.02} \kev$) and optically thick ($\tau = 12.7^{+0.5}_{-0.4}$) corona. These values are marginally consistent with that of obs3. In the case of obs2, we could not constrain the optical depth ($\tau$). The measured warm-corona radius ($\rm r_{\rm corona}$) remains nearly consistent between observations with values of $11.8^{+2.5}_{-1.8} \rm r_{g}$ and $5.9^{+1.9}_{-0.5} \rm r_{g}$ for \xmm{} and \suzaku{} observations respectively and indicates a compact corona. We note that the {\it optxagnf} model, similar to the {\it relxill}, favours a maximally rotating black hole spin to describe the broadband spectra. We could not constrain the spin parameter in {\it optxagnf} and got a lower limit of $>0.98$. We argue that a highly rotating black hole is required to model the broadband data in both ionized reflection and thermal Comptonization scenario and the presence of broad Fe emission line in the spectra further supports the idea of a highly spinning black hole at the centre of the AGN. The $\rm f_{pl}$ parameter in the {\it optxagnf} model determines the fraction of power below the coronal radius emitted in the hard comptonisation component. This parameter value is marginally consistent between the three observations with moderate values of $\sim 0.6$ and $\sim 0.4$ for \xmm{} and \suzaku{} observations respectively. This implies, in obs1, around 60 per cent of the gravitational energy released below $12 \rm r_{g}$, is emitted as the primary power-law emission with a photon index $\sim 1.74$ and the rest would help produce the soft excess. These values are consistent with recent studies found in other local Seyfert 1s \citep{2020MNRAS.tmp.2348G, 2019A&A...623A..11P}. Hence we conclude that for ESO~511--G030, highly ionized, relativistically blurred reflection from an untruncated accretion disk and the intrinsic thermal Comptonization of disk seed photons by a warm and optically thick compact corona, both combined with the neutral reflection from the torus, provide a good description to the multi-epoch optical-UV to hard X-ray spectra with parameter values consistent with other Seyfert 1s.

\section{Conclusions}\label{sec:conclusions}

We have extensively studied the broadband optical-UV to hard X-ray spectra of the bare Seyfert 1 galaxy ESO~511--G030 using \xmm{} and \suzaku{} multi-epoch observations and investigated the spectral features observed in the source with a physically motivated set of models. We list the main conclusions below.

\begin{itemize}

	\item{The optical-UV to hard X-ray spectra of ESO~511--G030 is typical of local Seyfert galaxies and consists of four components. A UV bump with an inner radius temperature of $\sim 2-3 \ev$,  a power-law continuum with a photon index varying between $\Gamma =1.7 - 2.0$, a prominent soft-excess and a relatively broad Fe line emission at $\sim 6.4\kev$.}

	\item {We found that the source is accreting at a sub-Eddington rate ($\lambdaedd$ varies within $0.004-0.008$) between 2007 and 2012.}

	\item{ The soft X-ray excess in ESO~511--G030 can be described either as the thermal Comptonization of disk seed photons by a warm ($0.40^{+0.02}_{-0.02}\kev$), optically thick ($\tau = 12.7^{+0.5}_{-0.4}$) and compact ($< 15 \rm r_{g}$) corona or as the blurred reflection from an untruncated and moderate to highly ionized accretion disk. However, for the blurred reflection, the model requires some extreme configuration (e.g., $\rm A_{Fe}<0.7$ and $\rm R\sim 5$) of the disk and corona. }

	\item{We confirm the presence of a broad Fe emission line at $\sim 6.4\kev$ in the source spectra with an equivalent width of $62-95 \ev$.}

	\item {Our broadband spectral modelling of ESO~511-G030 with both sets of physical models favour a rapidly spinning black hole ($a>0.78$) and a compact corona, indicating a relativistic origin of the Fe emission line.}

\end{itemize}

\section{Acknowledgements}

The authors are grateful to the anonymous referee for insightful comments which improved the quality of the paper. RG acknowledges the financial support from Visva-Bharati University and IUCAA visitor programme. 

\section{Data availability:}

This research has made use of archival data of \suzaku{} and \xmm{} observatories through the High Energy Astrophysics Science Archive Research Center Online Service, provided by the NASA Goddard Space Flight Center. 

\software{HEAsoft (v6.27.2; HEASARC 2014), 
         SAS (v18.0.0; \citet{2004ASPC..314..759G}), 
         XSPEC (v12.11.0m; \citet{1996aspc..101...17a})}

\begin{table*}

{\footnotesize
\centering
  \caption{The X-ray observations of ESO~511--G030. \label{Table:obs}}
  \begin{tabular}{cccccccc} \hline\hline 

X-ray		& observation	&Short	&Date of obs	& Net exposure	& Net counts\\
Satellite	&id		&id	&		&	        &\\ \hline 

\xmm{}		&0502090201	&obs1	&05-08-2007	& $76\ks$	&1.16E+06\\
		
\suzaku{}	&707023020	&obs2	&22-07-2012	&$224\ks$	&3.18E+06\\ 
		&707023030	&obs3	&17-08-2012	&$52\ks$	&1.12E+06\\

\hline 
\end{tabular}  


}
\end{table*}


 \begin{table*}
 \centering
   \caption{The comparison of different models used to fit the excess around $6\kev$ for the joint analysis of two \suzaku{} and \xmm{} observations of ESO~511--G030.\label{Table:above3kev}
 \label{model_compare}} 
  \begin{tabular}{|c|c|c|c|c|} \hline 
 Models                                 & Obs-1     & Obs-2     &Obs-3      & Simultaneous fit \\\hline
 Used					& $\cd$	    & $\cd$     & $\cd$     & $\cd$             \\\hline

 model 1 $+$ zxipcf                     &209/109    &1091/883   &567/437    & 1868/1420 \\
 model 1 $+$ narrow Gaussian		&145/105    &945/881    &513/433    & 1603/1417\\	  
 model 1 $+$ broad Gaussian		&131/104    &942/880    &510/433    & 1587/1414\\
 model 1 $+$ diskline			&133/104    &943/879	&512/433    & 1587/1414 \\
 model 1 $+$ relline			&137/105    &946/880    &516/433    & 1599/1411\\	  

   \hline
 \end{tabular} \\ 
 {Notes: Model 1 is above $3\kev$ powerlaw fit modified by the Galactic absorption.} \\
 {The normalization parameter of all the models were made free for each observations.}
 \end{table*}

\begin{table*}

\centering

  \caption{The best fit parameters of the baseline phenomenological models for the \xmm{} and \suzaku{} observations of ESO~511--G030.  \label{Table:pheno}}
{\renewcommand{\arraystretch}{1.5}
\setlength{\tabcolsep}{1.5pt}
\begin{tabular}{cccccccc} \hline\hline
	
Models 		& Parameter 				& obs1 		        & obs2  		   & obs3 		\\  \hline 

Gal. abs.  	& $\nh \,(\times 10^{20}\, \cmsqi)$ 	& $ 4.34$ (f)     	& $ 4.34$ (f)     	   & $ 4.34$ (f)     	\\


 powerlaw 	& $\Gamma$         			& $1.77^{+0.03}_{-0.02}$ & $1.80^{+0.01}_{-0.01}$  & $1.96^{+0.01}_{-0.01}$ \\
                & norm ($10^{-3}$) 			& $5.13^{+0.01}_{-0.01}$ & $5.21^{+0.01}_{-0.01}$  & $9.1^{+0.41}_{-0.46}$ \\

diskbb (1)  	& $T_{in}$ (keV)   			&$0.36^{+0.01}_{-0.02}$ &$0.23^{+0.01}_{-0.01}$    & $0.23$ (t)  \\
          	& norm ($10^{3}$) 			&$16^{+2}_{-3}$         & $2^{+1}_{-1}$            & $2$ (t)  \\

diskbb (2) 	& $T_{in}$ (keV)   			&$0.11^{+0.01}_{-0.01}$ & $0.07^{+0.01}_{-0.01}$   & $0.08^{+0.01}_{-0.01}$ \\
          	& norm ($10^{3}$) 			&$4.9^{+0.4}_{-0.4}$    & $97.5^{+1.4}_{-3.4}$     & $87.8^{+2.6}_{-0.6}$ \\

diskline  	& E($\kev$)        			& $6.31^{+0.03}_{-0.03}$ & $6.28^{+0.01}_{-0.02}$ & $6.28^{+0.03}_{-0.03}$  \\
          	& $\beta$          			& $-0.8^{+9.1}_{-0.7}$   & $9.9^{+0.1}_{-10.3}$   & $9.9$ (t)  \\
          	& $ \rm r_{in} \rm (r_{g})$             & $6 $ (f)               & $6 $ (f)               & $6 $ (f)                 \\
          	& Incl in($^\circ$)			& $10^{+4}_{-7}$         & $ 10$ (t)              & $ 10$ (t)                \\
          	& norm ($10^{-5}$) 			& $1.75^{+0.48}_{-0.45}$ & $1.54^{+0.15}_{-0.16}$ & $1.55^{+0.33}_{-0.33}$  \\


		&$\rm ^A$$\rm \dc/dof$			&$178/3$		  & $373/3$		   &$96/3$		    \\
Pexrav $^{B}$           & R              		& $-0.56$ (t)             &$-0.56^{+0.21}_{-0.12}$ & $-0.74^{+0.13}_{-0.15}$ 		       \\
			&$\rm ^A$$\rm \dc/dof$          & $12/1$                  & $34/2$                 & $26/1$ 		        \\\hline

        	 
$\rm reduced \cd$     &                  		& $ 1.28/168 $            & $1.13/1699 $            & $1.11/1108 $           \\\hline
\end{tabular}  

{$\rm ^A$ The $\dc$ improvement in statistics upon addition of the corresponding discrete component.\\
(f)  indicates a frozen parameter and (t) indicates parameters are tied between observations}\\

}

\end{table*}


\begin{table*}
\footnotesize
\centering
  \caption{The best-fitting parameters when we modelled the full X-ray energy band of \xmm{} and \suzaku{} observations of ESO~511--G030 with the ionized reflection model, {\it relxill} (Model 1) and the thermal Comptonization model, {\it optxagnf} (Model 2). \label{Table:X-ray}}

  \begin{tabular}{ccccccccccc} \hline
Component  & parameter                	     &            & obs1&    	                      &            & obs2&          &           &obs3&	\\\hline
	   & 				     & Model 1	  &               &   Model 2         & Model 1    &     & Model 2  & Model 1   &    & Model 2 \\\hline

Gal. abs.  & $\nh (10^{20} cm^{-2})$  	     & $ 4.34$ (f) &              & $ 4.34$ (f)	      & $4.34$(f)  &     &$4.34$(f) & $ 4.34$ (f)&   & $ 4.34$ (f)\\

{\it relxill }  &  $\rm A_{Fe}$            & $0.7^{+0.2}_{-0.1}$          &&-& $<0.6$                        &&-&  $0.5$(t)&                   &-    \\
 
           	&  $\log\xi (\xiunit)$     & $2.29^{+0.02}_{-0.06}$       &&-& $3.20^{+0.01}_{-0.01}$        &&-&  $3.20$ (t)&                 &-    \\ 

           	& $ \Gamma $               & $2.05^{+0.01}_{-0.01}$       &&-& $1.74^{+0.01}_{-0.01}$        &&-&  $ 1.88^{+0.01}_{-0.01}$&    &-    \\

		& $\rm E_{cut} (\kev)$     & $>297$                       &&-& $388$(t)                      &&-&  $ 388$(t)              &    &-    \\

           	&  $n_{rel}(10^{-5})^a$    & $8.72^{+0.08}_{-0.15}$       &&-& $11.24^{+0.16}_{-0.38}$       &&-&  $ 2.69^{+2.31}_{-0.12}$&    &-    \\
	   
           	&   $ q1$                  & $7.3^{+0.3}_{-0.3}$          &&-& $4.6^{+0.3}_{-1.7}$           &&-&  $8.2^{+0.2}_{-0.9}$ &       &-    \\
 
		&   $ a$                   & $>0.98$                      &&-& $0.99$ (t)                    &&-&  $0.99$ (t)&                 &-    \\

           	&   $\rm R(refl frac) $        & $1.2^{+0.1}_{-0.1}$      &&-& $0.5^{+0.1}_{-0.3}$           &&-&  $5.3^{+0.3}_{-0.3}$&        &-    \\

           	&   $\rm  R_{in}(r_{g})$       & $1.24$(f)                &&-& $1.24$(f)                     &&-&  $1.24$(f) &                 &-    \\

		&   $\rm  R_{br}(r_{g})$       & $4.9^{+0.4}_{-0.3}$      &&-& $4.3^{+0.2}_{-0.1}$           &&-&  $3.6^{+0.2}_{-0.3}$ &       &-    \\

           	&   $\rm  R_{out}(r_{g})$      & $400$ (f)                &&-& $400$(f)                      &&-&  $400$(f)&                   &-     \\

           	&   $i(degree) $           & $27^{+3}_{-1}$               &&-& $ 27$(t)                      &&-&  $27 $(t)&                   &-     \\ \hline

{\it MYTorusL}  &   $i(degree) $           & $45$ (t)&&          $45$ (t)       & $45$(f)               &    &$45$(f)             & $45$ (t)  &   &$45 $(t)        \\
		&  norm ($10^{-3}$)        & $7.37$(t)&&        $8.18$(t)       & $7.37^{+0.40}_{-0.37}$&  &$8.18^{+0.06}_{-0.04}$ & $7.37$ (t)&  & $8.18$(t)        \\

{\it MYTorusS } &  NH($10^{24} cm^{-2}$)   & $10.0$(t)&&        $10.0$ (t)     & $10$(f)                &    &$10$(f)               &$10.0$(t)&   &$10.0$ (t)       \\
		&  norm ($10^{-3}$)        & $3.20$(t)&&        $6.05$(t)      &$3.20^{+1.18}_{-0.71}$&     &$6.05^{+0.58}_{-0.39}$&$3.20$(t) &   &$6.05$(t)  \\\hline

optxagnf   & $ M_{BH}^b$            &-&&$4.57$(f)                                &-&&$4.57$(f)                 &-&&$4.57$(f)           \\
           & $d{\rm~(Mpc}) $        &-&&$95 $(f)                                 &-&&$95$(f)                   &-&&$95$(f)            \\
           &  $(\frac{L}{L_{E}})$   &-&&$0.004^{+0.001}_{-0.001}$                &-&&$0.008^{+0.001}_{-0.001}$ &-&&$0.008^{+0.001}_{-0.001}$                \\ 
           &  $ kT_{e} (\kev)$      &-&&$0.39^{+0.12}_{-0.02}$                   &-&&$0.10^{+0.15}_{-0.04}$    &-&&$0.54^{+0.02}_{-0.10}$            \\ 
           &  $ \tau $              &-&&$12.9^{+1.7}_{-1.9}$                      &-&&$>11.8$     	       &-&&$7.9^{+3.0}_{-2.5}$(t)              \\
           &  $\rm  r_{cor}(r_{g})$ &-&&$11.8^{+4.7}_{-3.3}$                     &-&&$5.2^{+2.7}_{-0.9}$       &-&&$5.2$(t)             \\
           &  $ a $                 &-&&$>0.78$                                  &-&&$ 0.99$(t)                &-&&$ 0.99$(t)            \\
           &  $\rm  f_{pl}$         &-&&$0.56^{+0.03}_{-0.05}$                   &-&&$0.41^{+0.08}_{-0.04}$    &-&&$0.59^{+0.02}_{-0.07}$   \\
           &  $ \Gamma $            &-&&$1.74^{+0.01}_{-0.02}$                   &-&&$1.79^{+0.02}_{-0.02}$    &-&&$1.86^{+0.01}_{-0.03}$             \\\hline

	   & $\rm reduced \cd $     & $1.37/170$    && $1.18/171$      & $1.13/1703$    && $1.14/1706$   & $ 1.05/1110$        &&   $ 1.09/1112$                  \\\hline 
\end{tabular} 
{Notes: Model 1 = {\sc tbabs $\times$ (relxill +MyTorus)}; Model 2 = {\sc tbabs $\times$ (optxagnf +MyTorus)}\\
        The {\it MYTorus} parameters for obs1 and obs3 are tied with obs2 which is the longest \suzaku{} observation (See Table~\ref{Table:obs})\\
	(f) indicates a frozen parameter. (*) indicates parameters are not constrained.\\
	(a) $n_{rel}$ reperesent normalization for the model {\it relxill}\\
        (b) in units of $10^{8}\msol$.}
\end{table*}

\begin{table*}
\footnotesize
\centering
  \caption{The best-fitting parameters when we modelled the optical-UV to hard X-ray energy band of \xmm{} and \suzaku{} observations of ESO~511--G030 with the ionized reflection model, {\it relxill} (Model 1) and the thermal Comptonization model, {\it optxagnf} (Model 2). \label{Table:sed}}

  \begin{tabular}{ccccccccccc} \hline
Component  & parameter                	     &            & obs1&    	                      &            & obs2&          &           &obs3&	\\\hline
	   & 				     & Model 1	  &               &   Model 2         & Model 1    &     & Model 2  & Model 1   &    & Model 2 \\\hline

Gal. abs.  & $\nh (10^{20} cm^{-2})$  	     & $ 4.34$ (f) &              & $ 4.34$ (f)	      & $4.34$(f)  &     &$4.34$(f) & $ 4.34$ (f)&   & $ 4.34$ (f)\\

diskbb    &  $T_{in}(\kev)$                & $0.004^{+0.001}_{-0.001}$&&-& $0.004$(t)                &&-& $0.004$(t)  &      &\\

          &  norm ($10^{10}$)              & $2.54^{+2.18}_{-1.42}$   &&-& $2.54$(t)                 &&-& $2.54$(t) &       &      \\\hline

{\it relxill }  &  $\rm A_{Fe}$            & $0.7$ (f)         &&-& $0.5$ (f)                        &&-&  $0.5$ (f)&                   &-    \\
 
           	&  $\log\xi (\xiunit)$     & $2.29$ (f)        &&-& $3.20$(f)                        &&-&  $3.20$ (f) &                 &-    \\ 

           	& $ \Gamma $               & $2.05$ (f)        &&-& $1.74$(f)                        &&-&  $1.88$ (f) &    &-    \\

		& $\rm E_{cut} (\kev)$     & $388$ (f)         &&-& $388$(t)                      &&-&  $ 397$(t)              &    &-    \\

           	&  $n_{rel}(10^{-5})^a$    & $8.27$(f)         &&-& $11.24$ (f)                   &&-&  $2.69$(f) &    &-    \\
	   
           	&   $ q1$                  & $7.3$ (f)         &&-& $4.6$ (f)                     &&-&  $8.2$ (f) &       &-    \\
 
		&   $ a$                   & $0.99$(f)         &&-& $0.99$ (t)                    &&-&  $0.99$ (t)&                 &-    \\

           	&   $\rm R(refl frac) $        & $1.2$(f)      &&-& $0.5$ (f)                     &&-&  $5.3$ (f) &                     &-    \\

           	&   $\rm  R_{in}(r_{g})$       & $1.24$(f)     &&-& $1.24$(f)                     &&-&  $1.24$(f) &                 &-    \\

		&   $\rm  R_{br}(r_{g})$       & $4.9$ (f)     &&-& $4.3$ (f)                     &&-&  $3.6$ (f) &       &-    \\

           	&   $\rm  R_{out}(r_{g})$      & $400$ (f)     &&-& $400$(f)                      &&-&  $400$(f)&                   &-     \\

           	&   $i(degree) $           & $27$ (f)          &&-& $ 27$(t)                      &&-&  $27 $(t)&                   &-     \\ \hline

{\it MYTorusL}  &   $i(degree) $           & $45$ (t)&&          $45$ (t)       & $45$ (f)        &&  $45$(f)     & $45$ (t)  &   &$45$ (t)           \\
		&  norm ($10^{-3}$)        & $7.37$(t)&&        $8.18$(t)      & $7.37$(f)       &&  $8.18$(f)  & $7.37$ (t)&   & $8.18$(t)        \\

{\it MYTorusS } &  NH($10^{24} cm^{-2}$)   & $10.0$ (t)&&        $10.0$ (t)     & $10.0$(f)       &&   $10.0$(f)  &$10.0$(t)  &   &$10.0$ (t)       \\
		&  norm ($10^{-3}$)        & $3.20$(t)&&        $6.05$(t)      &$3.20$ (f)       &&   $6.05$(f) &$3.20$(t) &   &$6.05$(t)  \\\hline

optxagnf   & $ M_{BH}^b$            &-&&$4.57$(f)                                &-&&$4.57$(f)                 &-&&$4.57$(f)           \\
           & $d{\rm~(Mpc}) $        &-&&$95 $(f)                                 &-&&$95$(f)                   &-&&$95$(f)            \\
           &  $(\frac{L}{L_{E}})$   &-&&$0.004^{+0.001}_{-0.001}$                &-&&$0.007^{+0.001}_{-0.001}$ &-&&$0.008^{+0.001}_{-0.001}$                \\ 
           &  $ kT_{e} (\kev)$      &-&&$0.40^{+0.02}_{-0.02}$                   &-&&$0.09^{+0.10}_{-0.03}$    &-&&$0.53^{+0.03}_{-0.10}$      \\ 
           &  $ \tau $              &-&&$12.7^{+0.5}_{-0.4}$                     &-&&$>11.9$                   &-&&$7.9^{+2.0}_{-1.7}$              \\
           &  $\rm  r_{cor}(r_{g})$ &-&&$11.8^{+2.5}_{-1.8}$                     &-&&$5.9^{+1.9}_{-0.5}$       &-&&$5.9$(t)    \\
           &  $ a $                 &-&&$>0.98$                                  &-&&$ 0.99$(t)                &-&&$ 0.99$(t)            \\
           &  \rm $ f_{pl}$         &-&&$0.57^{+0.02}_{-0.02}$                   &-&&$0.41^{+0.06}_{-0.08}$    &-&&$0.60^{+0.06}_{-0.06}$   \\
           &  $ \Gamma $            &-&&$1.74^{+0.02}_{-0.02}$                   &-&&$1.79^{+0.01}_{-0.01}$    &-&&$1.86^{+0.01}_{-0.01}$             \\\hline

	   & $\rm Reduced \cd $     & $1.38/170$    && $1.38/171$     & $1.13/1704$    && $1.14/1706$   & $ 1.05/1108$        &&   $ 1.09/1112$                  \\\hline 
\end{tabular} 
{Notes: Model 1 = {\sc tbabs $\times$ (diskbb + relxill + MyTorus)}; Model 2 = {\sc tbabs $\times$ (optxagnf +MyTorus)}\\
	The {\it MYTorus} parameters for obs1 and obs3 are tied with obs2 which is the longest \suzaku{} observation (See Table~\ref{Table:obs})\\
        (f) indicates a frozen parameter. (*) indicates parameters are not constrained.\\
	(a) $n_{rel}$ reperesent normalization for the model {\it relxill}\\
        (b) in units of $10^{8}\msol$.}

\end{table*}


\bibliographystyle{aasjournal}
\bibliography{mybib}

\end{document}